\documentclass{evn2002}
\usepackage{graphicx}

\begin{document}
   \title{Kinematics of parsec-scale structures in AGN:\\ the 2cm VLBA Survey
          }

   \author{E. Ros\inst{1}
          \and
          K. I. Kellermann\inst{2}
          \and 
          M. L. Lister\inst{2}
          \and
          J. A. Zensus\inst{1}
          \and
          M. H. Cohen\inst{3}
          \and
          R. C. Vermeulen\inst{4}
          \and
          M. Kadler\inst{1}     
          \and
          D. C. Homan\inst{2}  
          }

   \institute{Max-Planck-Institut f\"ur Radioastronomie, Auf dem H\"ugel 69,
              D-53121 Bonn, Germany 
         \and
             National Radio Astronomy Observatory, 520 Edgemont Rd.,
             Charlottesville, VA 22903, US    
         \and
             California Institute of Technology, Department of Astronomy,
             MS 105-24, Pasadena, CA 91125, US
         \and
             Netherlands Foundation for Research in Astronomy, Postbus 2,
             7990 AA Dwingeloo, The Netherlands       
             }

   \abstract{
We are investigating the kinematics of jets in active galactic nuclei 
on parsec scales by studying a representative
population of sources.  This study is being carried out using the
Very Long Baseline Array at 15\,GHz, with more than 800 images taken since
1994. In this contribution we present an overview of the diversity
of kinematics
for a complete sample of sources.  
   }

   \maketitle
%


Since 1995, we have been performing a monitoring program on
a sample of over 170 Active Galactic Nuclei (AGN)
using the VLBA\footnote{The Very Long Baseline Array of the U.S.\
National Radio Astronomy Observatory is operated by
Associated Universities, Inc., under cooperative agreement with the
U.S.\ National Science Foundation.} at 15\,GHz to study the 
jet bending, pattern motions, accelerations or changes in
the component strength, morphology and Lorentz factors.


In 1994, we began by observing a sample of 132 compact AGN approximately every
6 months with the VLBA at 15\,GHz
(Kellermann et al.\ \cite{kel98}, herafter K98).
In 1997 we added 42 sources to our sample (Zensus et al.\ \cite{zen02}, 
hereafter Z02).
Approximately 40 observing sessions have been carried out since August 1994 
with the VLBA\footnote{Programs BZ4, BZ14, BK37, 
BZ22, BK52, BK68 and BK77.  
The next observations, will have the code BL111, corresponding to
a large proposal named MOJAVE (Monitoring of Jets in AGN with VLBA 
Experiments).}, giving one observation every 6--18 months for
each source.  

To make a robust statistical analysis of the kinematics and source
properties, we need to define a ``complete'' sample.  The initial criteria 
we followed were: a flat spectrum above 500\,MHz and a 15\,GHz flux
density over 1.5\,Jy for objects
above the celestial equator and over 2\,Jy
for objects with declination
between 0\degr\ and $-20$\degr.  The catalogue from
Stickel et al.\ (\cite{sti94}), from which the sources were initially
chosen, is complete only at 5\,GHz, so we needed to identify all
the sources with the suitable flux density level from the available
flux density surveys\footnote{For instance,
the University of Michigan Radio Observatory (UMRAO) survey, Aller et al.\
in preparation; the RATAN-600, Kovalev et al.\ \cite{kov99};
or Mets\"ahovi, 
Ter\"astranta et al.\ (\cite{ter01}).}.  From the 174 available
sources (see K98 and Z02) we discarded the weak VLBI cores and selected
the sources with at least 4 VLBA epochs in our survey.  We rejected
the sources with final {\sc clean}ed or UMRAO flux densities below
the 1.5/2\,Jy level. That gives a total of 74 sources which represent
a ``complete" subsample of our survey sources.


After imaging the radio sources (see K98 and Z02) we measured
the positions of the absolute and relative peaks of brightness in
the images using the ${\cal A}{\cal I}{\cal P}{\cal S}$ task {\sc imfit}.
For some sources with very close components or other problems,
we performed a model fitting of the data using
the task {\sc modelfit} in {\sc difmap}.  
We cross-identified the same components 
at different epochs by using several criteria:
flux density evolution, similar relative positions
with respect to
the main component (the ``core"), etc.  
So far, we have reasonably well-determined motions for about 100 of
the 174 sources in our sample extending over a time baseline of
4 to 6 years.  

\begin{figure}[htbp]
\vspace{-5pt}
\centering
\includegraphics[bb=95 85 770 510,clip,width=0.44\textwidth]{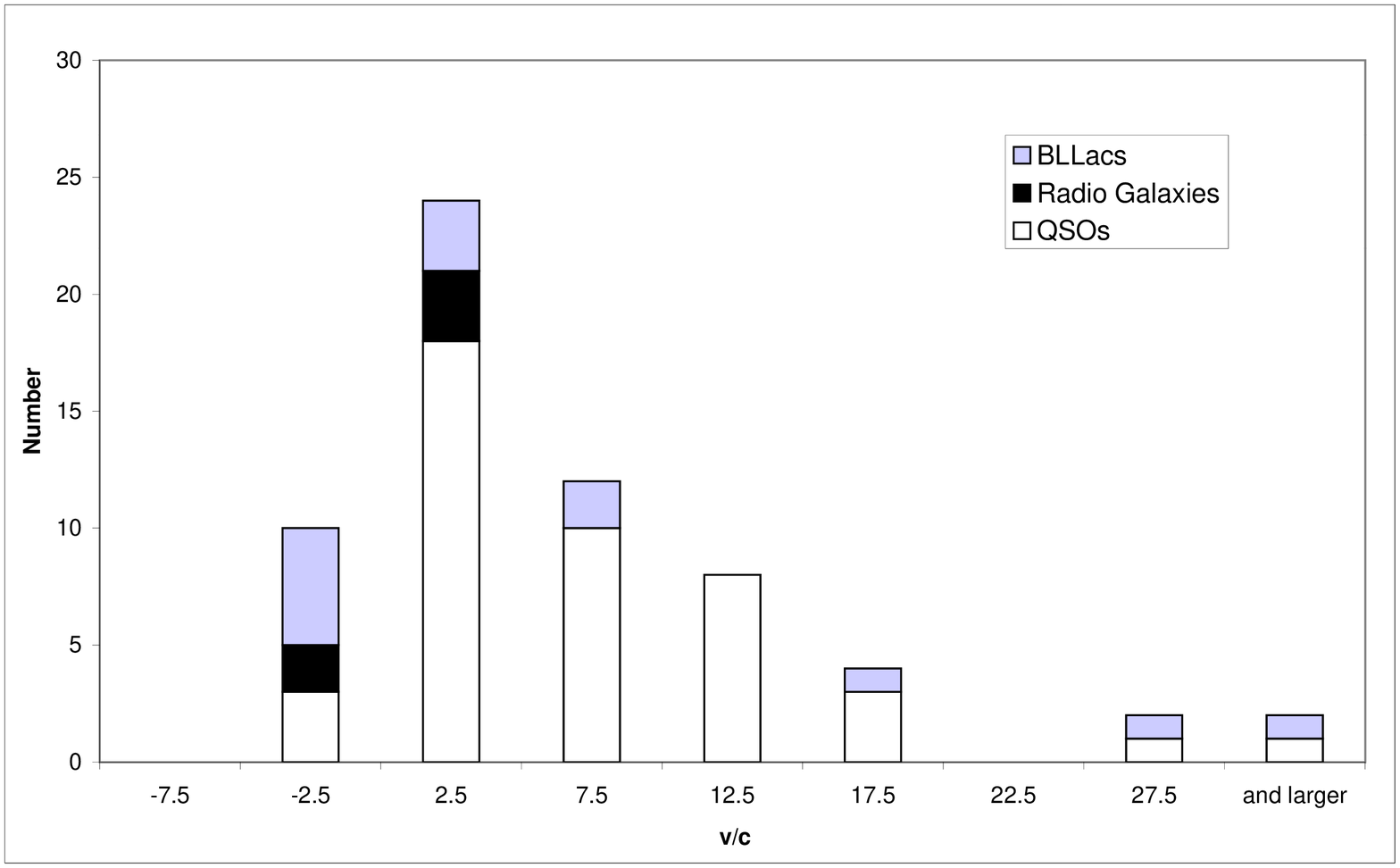}
\caption{Distribution of $\beta_\mathrm{app}$ for the different object
classes of the ``complete'' subsample of sources (see text).}
\label{fig:lorentz}
\end{figure}

\begin{figure}[bth]
\vspace{-8pt}
\includegraphics[clip,width=0.47\textwidth]{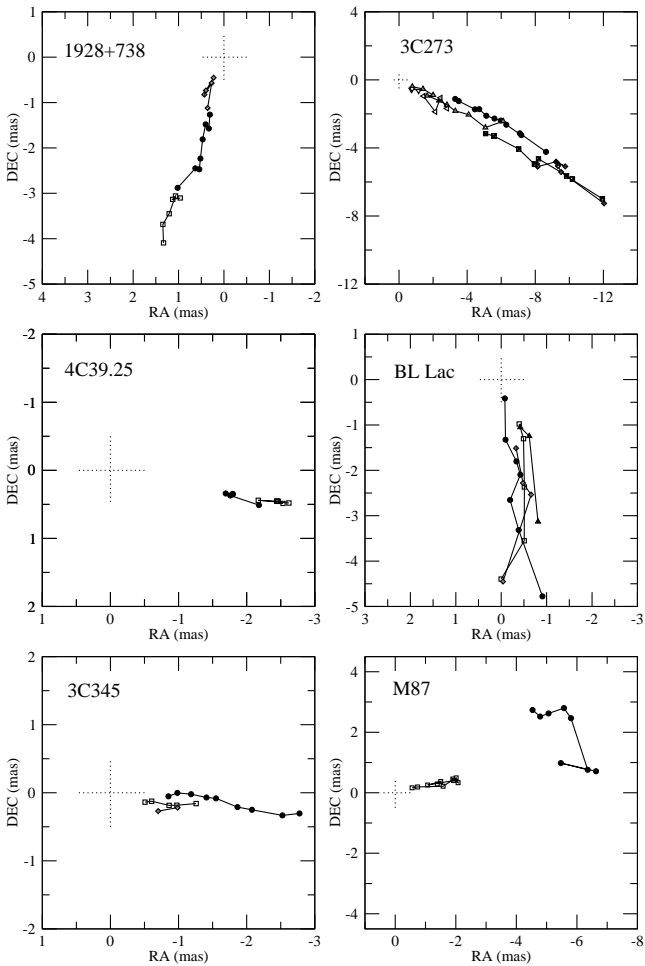}
\caption[]{
Apparent sky trajectories of jet components in the AGN 1928+738, 3C\,273, 
4C\,39.25, BL\,Lac, 3C\,345, and M87.
The crosses indicate the position of the VLBI cores.
\label{fig:xy}
}
\end{figure}

\begin{figure}[bth]
\vspace{-8pt}
\includegraphics[clip,width=0.47\textwidth]{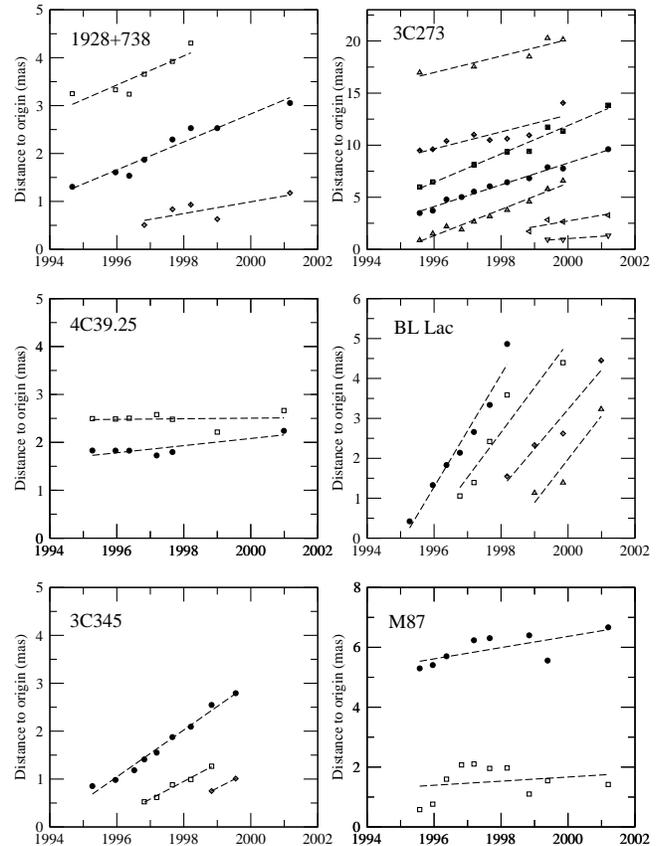}
\caption[]{
Angular separation $r$ versus time corresponding to the radio
sources and components from Fig.~\ref{fig:xy}.
\label{fig:rt}
}
\end{figure}

\begin{table*}[htbp]
\caption{Component speeds in the sources shown in Fig.~\ref{fig:xy} and 
\ref{fig:rt}, ordered by the component proximity to the core.\label{table:speeds}}
\begin{footnotesize}
\begin{tabular}{lcc|c@{~}c|c@{~}c|c@{~}c|c@{~}c|c@{~}c|c@{~}c|c@{~}c}
Source    & $z$ & Id.\ & 
$\dot{r}^\mathrm{\,(a)}$ & $\beta_\mathrm{app}$ &
$\dot{r}^\mathrm{\,(a)}$ & $\beta_\mathrm{app}$ &
$\dot{r}^\mathrm{\,(a)}$ & $\beta_\mathrm{app}$ &
$\dot{r}^\mathrm{\,(a)}$ & $\beta_\mathrm{app}$ &
$\dot{r}^\mathrm{\,(a)}$ & $\beta_\mathrm{app}$ &
$\dot{r}^\mathrm{\,(a)}$ & $\beta_\mathrm{app}$ &
$\dot{r}^\mathrm{\,(a)}$ & $\beta_\mathrm{app}$ \\
\noalign{\smallskip} \hline \noalign{\smallskip}
1928+738  & 0.300 & Q & 0.12    & 1.04    & 0.29  & 0.52 & 0.30 & 1.39  &      &       &       &       &      &      &      &    \\
3C\,273   & 0.160 & Q & 0.23    & 2.56    & 0.55  & 6.23 & 1.26 & 14.24 & 1.05 & 11.80 &  1.36 & 15.36 & 0.82 & 9.20 & 0.79 & 8.94  \\
4C\,39.25 & 0.700 & Q & 0.075   & 3.19    & 0.006 & 0.26 &      &       &      &       &       &       &      &      &      &  \\
BL\,Lac   & 0.070 & B & 0.99    &  5.0    & 1.09  & 5.5  & 1.13 & 5.7   & 1.41 & 7.1   &       &       &      &      &      & \\ 
3C\,345   & 0.590 & Q & 0.36    & 13.4    & 0.37  & 13.7 & 0.49 & 18.2  &      &       &       &       &      &      &      & \\
M87       & 0.004 & G & 0.07    & 0.02    & 0.19  & 0.05 &      &       &      &       &       &       &      &      &      & \\
\noalign{\smallskip} \hline \noalign{\smallskip}
\end{tabular}
~\\
$^\mathrm{a}$: Given in mas\,yr$^{-1}$\\
\end{footnotesize}
\end{table*}

From the 74 ``complete'' sources, we do have kinematical data
for 63 of these sources (14 BL\,Lac objects, 5 radio galaxies, and 44 QSOs).
In Fig.~\ref{fig:lorentz}, we present the 
distribution of apparent speeds\footnote{
For $H_0=65$\,km\,Mpc$^{-1}$\,s$^{-1}$\,
$\Omega_m=0.3$, and $\Omega_\Lambda$=0.7.
} for their brightest components.
The median of $\beta_\mathrm{app}$ for the different classes of objects is
of 1.84, 0.02, and 5.23 (BL, G and QSOs, respectively).  

As an example of our kinematical analysis, we show in 
Figs.~\ref{fig:xy} and \ref{fig:rt} a selection of the
component motions for six well-known AGN.  Fig.~\ref{fig:xy} presents
the sky trajectories of the components.
The curved paths in 1928+738 and 3C\,345 are remarkable.
The components in M87 are extended and therefore their positions are
not well-defined.  
BL\,Lac shows recurrent component trajectories.
In Fig.~\ref{fig:rt} we show the
time evolution of the component
distance to the core.  The slopes of the
linear regression fits are the speeds given in
Table~\ref{table:speeds} (their uncertainties, not given here,
are below 5\%).

A detailed description of the extensive
kinematic analysis on the radio sources
of the survey will be given in a forthcoming paper (Kellermann et
al., in preparation).

%


\end{document}